\documentclass[9pt,twocolumn,twoside]{opticajnl}
\journal{opticajournal} 
\usepackage{gensymb}
\setboolean{shortarticle}{true}
\usepackage{hyperref}

\usepackage{lineno}

\title{Young's double-slit experiment with anisotropic GHz surface acoustic waves on gallium arsenide}

\author[*,1]{Thomas Steenbergen}
\author{Matteo Fisicaro}
\author{Krystian Czerniak}
\author{Matthijs Rog}
\author{Kaveh Lahabi}
\author{Wolfgang L\"offler}

\affil[]{Leiden Insitute of Physics, Leiden University,
P.O. Box 9504, 2300 RA Leiden, The Netherlands}
\affil[*]{steenbergen@physics.leidenuniv.nl}
\doi{}

\begin{abstract}
We demonstrate Young’s double-slit experiment with GHz surface acoustic waves (SAWs) on gallium arsenide (GaAs). This experiment differs from the well-known optical case due to the anisotropy of SAW propagation on GaAs. We generate SAWs using an interdigital transducer (IDT), and the double-slit is realized by focused ion beam milled grooves which block the SAWs. We measure the out-of-plane SAW displacement with an optical scanning interferometer and apply a spatial Fourier filtering technique, enabling the observation of the two-slit interference pattern, as well as higher order diffraction and near field interference effects. We find reasonable agreement between the measured far-field interference pattern and anisotropic Huygens-Fresnel simulations based on previously reported angle-dependent SAW velocities on GaAs.
\end{abstract}

\setboolean{displaycopyright}{false} 

\begin{document}

\maketitle

Double-slit interference experiments have fascinated physicists for more than two centuries, revealing the fundamental duality of the wave and particle behavior of light and matter. The first experiment of this kind was performed in 1801 by Young \cite{Young1804}, in which he demonstrated the interference of light and consequently its wave-like nature. The first double-slit with single electrons was performed in 1974, demonstrating the statistical nature of the build-up of the interference pattern \cite{Merli1976}. Later, experiments were carried out with neutrons \cite{Zeilinger1988}, C60 and even larger molecules \cite{Arndt1999,Fein2019}, Bose-Einstein condensates \cite{Andrews1997} and quasi-particles including surface plasmons \cite{Zia2007}. For surface acoustic waves, a double-slit experiment is still missing, although other interference experiments have been explored, such as extraordinary transmission \cite{Mezil2016} and propagation in phononic crystals \cite{Ash2021}.

Surface acoustic waves (SAWs) are mechanical waves that propagate on the surface of a material and have found a variety of applications in modern technology, mainly because they have low loss and because they can be easily excited electrically using interdigital transducers (IDTs). Compared to electromagnetic waves, GHz-range SAWs have a much smaller micron-scale wavelength, which enabled miniaturization of GHz electronic components such as radio frequency (RF) filters. The strong confinement of SAWs to the surface makes them ideal as sensors in chemistry \cite{Go2017} and biology \cite{Rocha-Gaso2009}. In more recent developments, SAWs have been used in quantum physics research, enabled by their long coherence times at ultra-low temperatures and the ability to couple to a variety of quantum systems, such as superconducting qubits \cite{Sletten2019}, nitrogen vacancy centers in diamond \cite{Golter2016}, defects in two-dimensional materials \cite{Patel2024} and semiconductor quantum dots hosted in gallium arsenide (GaAs) \cite{Decrescent2022}. 

Most piezoelectric materials exhibit anisotropic SAW velocities, including GaAs with a zinc-blende lattice structure. To our knowledge, the effect of anisotropic wave velocities on double-slit interference has to date only been theoretically studied for 2D electron systems \cite{araujo}, where a narrowing of the double-slit interference pattern and beam steering was found. 

Here we show Young's double-slit experiment with SAWs on GaAs, by launching SAWs from an IDT onto a double slit, which is fabricated using a focused ion beam (FIB). By measuring the complex SAW field with sub-$\mu$m resolution over the full device and after applying a spatial Fourier filtering technique, we reveal the double-slit interference pattern and validate both beam steering and the reshaping of the interference pattern by the anisotropic wave velocities.



\begin{figure}[ht]
\centering
\includegraphics[width=1\linewidth]{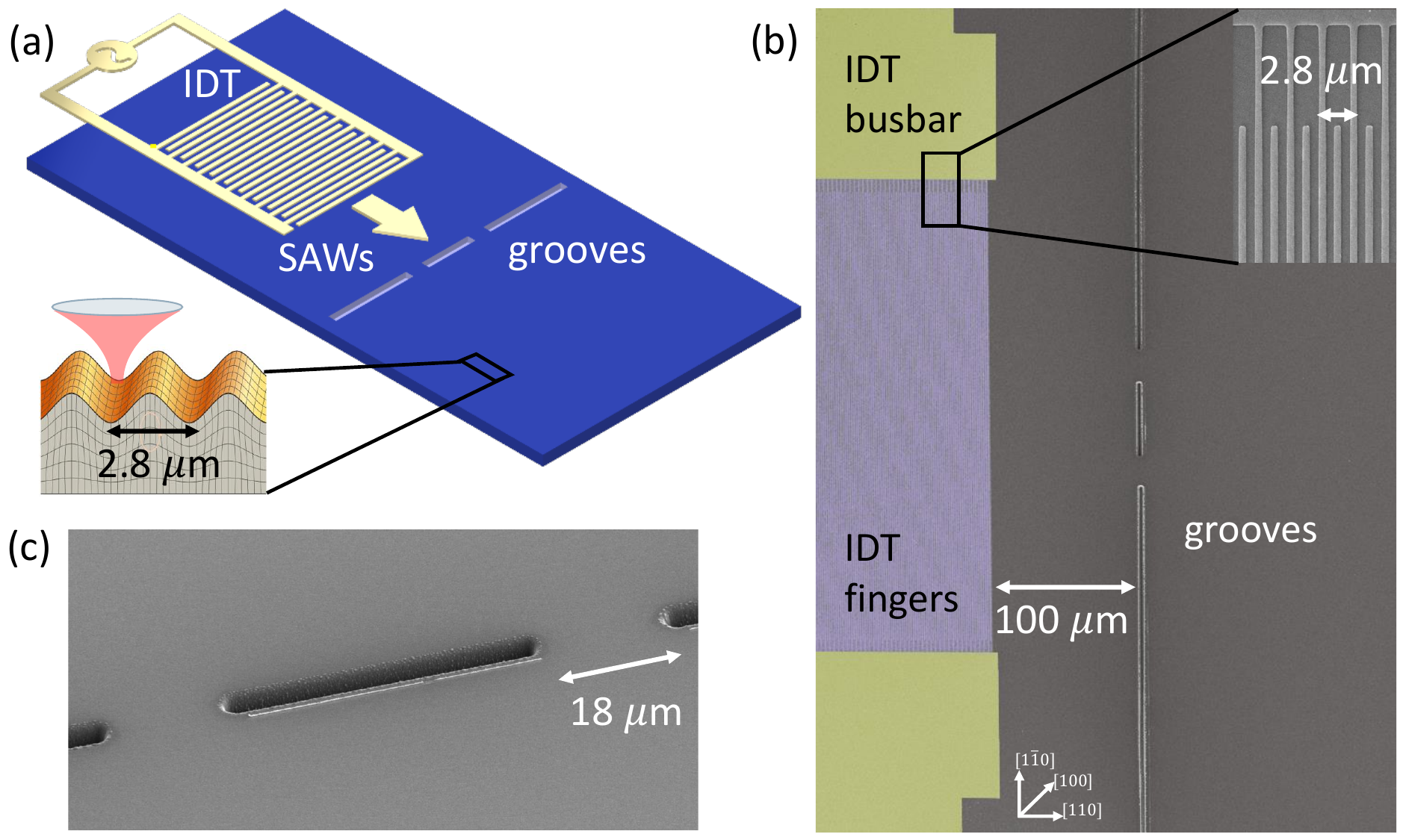}
\caption{Experimental setup. Schematic of the double-slit experiment (a) with inset showing focused light on the sample for optical SAW measurement. False-color scanning electron microscope (SEM) image of the IDT and grooves (b), where the inset shows the IDT fingers. Close-up SEM image of the grooves (c), at an angle of 52$\degree$.}
\label{fig:exp_setup}
\end{figure}

In our device, the SAWs are generated by an IDT and the double-slit is realized by grooves which partially block the SAWs, see Fig.\ref{fig:exp_setup}(a). The IDT is fabricated on a (001)-cut GaAs substrate using an electron-beam lithography lift-off process and consists of a 5 nm chromium adhesion layer and a 30 nm-thick gold layer. The IDT has 50 finger pairs and the fingers are 700 nm wide and 310 $\mu$m long. The IDT periodicity is chosen to be 2.8 $\mu$m [see inset Fig. \ref{fig:exp_setup}(b)], to efficiently excite the SAWs at 996 MHz ($v_{SAW} \approx 2860$ m/s in the [110] direction). When an RF voltage is applied to the opposing combs of the IDT, the oscillating electric field leads via the inverse piezoelectric effect to the excitation of SAWs in the [110] direction. To form the double slit, three grooves are milled in the substrate by a FIB with a depth of at least 3 $\mu$m, as measured by a profilometer. The angle between the grooves and the IDT fingers is less than 0.1$\degree$. After the FIB process, some debris is visible in the optical microscope images, and we find that the DC resistance between opposing IDT combs is only 400 k$\Omega$, while usually being around 1 G$\Omega$. Although this finite resistance slightly degrades the IDT performance, it does not pose a serious limitation.

The out-of-plane SAW displacement is measured with an optical scanning Michelson interferometer, in which laser light is focused on the SAW sample with a sufficiently small spot size of around 2.8 $\mu$m. The optical path length is modified by the out-of-plane component of the SAW displacement leading to a GHz modulation of the interferometric signal that is read-out using a lock-in technique, allowing for the extraction of the amplitude and phase of the SAW displacement. In order to obtain high-resolution spatial scans, the double-slit sample is mounted on a motorized XYZ-translation stage with a step size of 14 nm in the direction of the SAW propagation. For a more detailed description of the interferometric measurement setup see Ref. \cite{Fisicaro2025}.

\begin{figure}[H]
\centering
\includegraphics[width=1\linewidth]{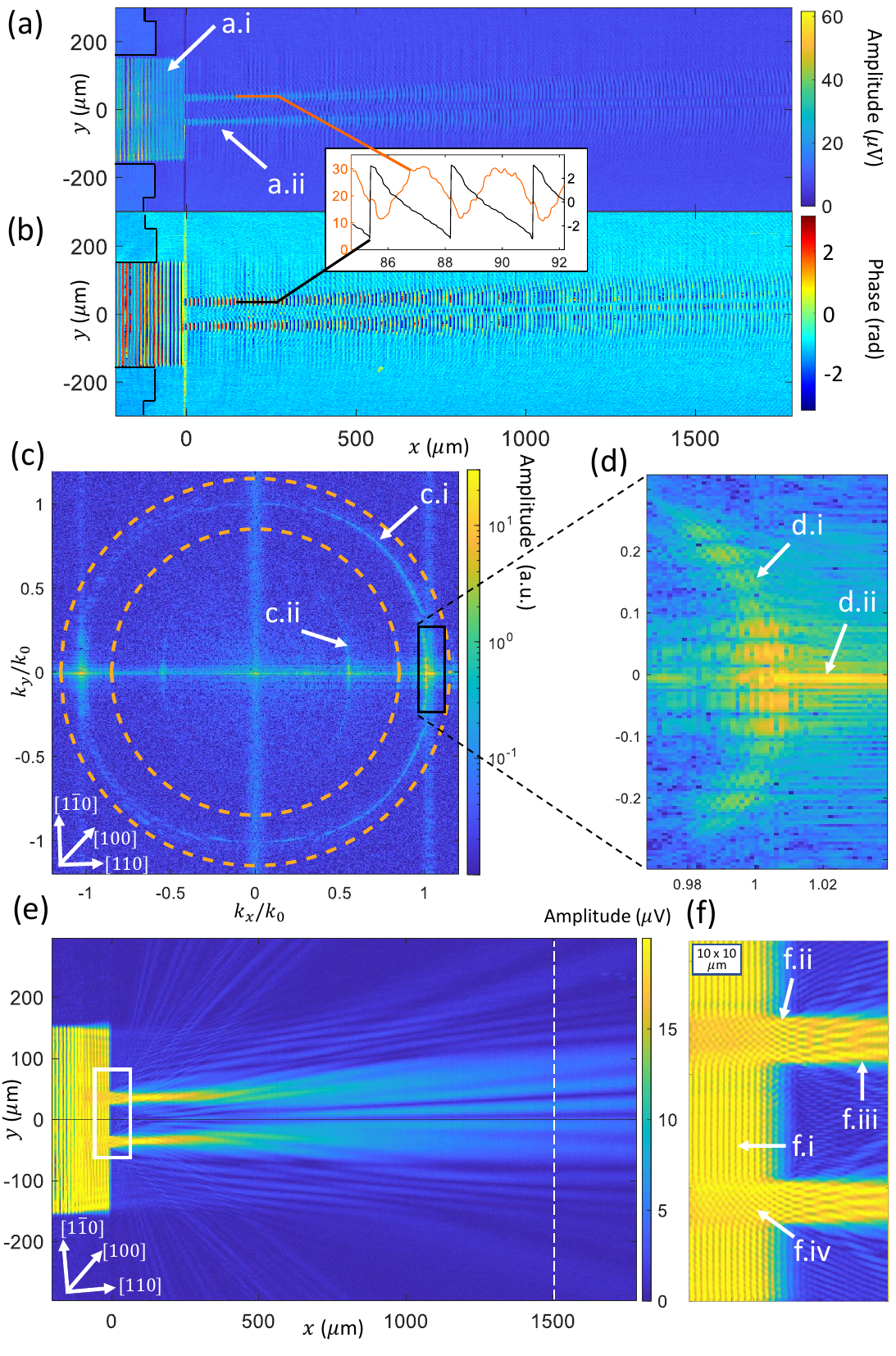}
\caption{Raw data, spatial FFT and filtered measurements. Rms amplitude (a) and phase (b) of the 996 MHz demodulated interferometric signal, where the black lines indicate the IDT busbars. Absolute value of the spatial FFT (c) of the complex measurement, where the orange dashed circles indicate the standard deviation $\sigma$ of the Gaussian window function for filtering. Zoom-in (d) of the FFT (same color scale). Filtered measurement of the SAW amplitude (e) and zoom-in near the slits (f), with the same saturated color scale as in (e) for increased visibility in the far field.}
\label{fig:Raw_data}
\end{figure}

First, we show experimental amplitude and phase measurements of the SAW displacement in Fig. \ref{fig:Raw_data}(a) and (b), respectively. Between the IDT and the grooves we observe a very strong signal (a.i), which is due to SAW confinement. Behind the slits, SAWs travel in the positive $x$-direction (a.ii) in two beams, indicating that the grooves block the SAWs efficiently. Surprisingly, the amplitude measurement of the transmitted signal exhibits oscillations [see inset] with a periodicity of $\lambda_{SAW} \approx$ 2.87 $\mu$m, which is the expected SAW wavelength in the [110] direction at 996 MHz. This signal is relatively weak compared to the background, which we attribute to residual RF crosstalk and possibly to the presence of bulk acoustic waves (BAWs). Interference between this background signal and the SAW signal results in spatial oscillations in amplitude measurements. We now show that these oscillations, as well as the background signal, can be removed efficiently using a spatial Fourier filtering technique.

We combine the amplitude and phase information in a complex field, and take the spatial 2D (fast) Fourier transform (FFT), of which the absolute value is shown in Fig. \ref{fig:Raw_data}(c) on a logarithmic scale. The axes are oriented such that a signal at $(k_x,k_y)$ corresponds to a spatial periodicity in the direction of $(x,y)$ in Fig. \ref{fig:exp_setup}(a,b), normalized to the expected wavenumber $k_0=2\pi/\lambda_{SAW}$. The FFT amplitude quantifies the strength of the signal in this direction. In this FFT, we observe a circular signal (c.i) around $|k|=\sqrt{k_x^2+k_y^2}\approx 2\pi/\lambda_{SAW}$. This indicates SAWs traveling in all directions, but predominantly in the positive $x$ direction. The signal at negative $k_x$ is caused by reflections from the grooves and standing waves inside the IDT. We also observe a signal at a wavelength of 5.3 $\mu$m (c.ii), corresponding to a velocity of 5240 m/s in the [110] direction, which we attribute to a longitudinal bulk mode. 

A close-up of the signal related to SAWs propagating in the positive $x$-direction is shown in Fig. \ref{fig:Raw_data}(d). Here, we observe an arc (d.i) corresponding to SAW propagation over the free GaAs surface, exhibiting the double-slit interference fringes in the Fourier domain and a vertical asymmetry, which indicates a slight tilt between the [110] crystal axis and the measurement frame. We also observe a signal at slightly lower wavelengths, which is more spread out in the $k_x$ direction (d.ii). This signal is related to SAWs propagating over metalized structures (the IDT) mostly parallel to the $x$-axis. The difference in wavelength between (d.i) and (d.ii) is the result of mass loading by the IDT fingers, lowering the SAW velocity and consequently the wavelength for SAWs propagating over the IDT. This is verified with selective spatial FFTs of the IDT region and free surface regions.

Now, we filter the raw data using a Fourier filtering technique to remove most of the non-SAW related contributions from the measured signal. To do this, we apply a circular Gaussian window function to the amplitude of the Fourier-transformed data, centered around $|k|= 2\pi / \lambda_{\mathrm{SAW}}$ with a standard deviation $\sigma = 15\%$ of the central $|k|$ value (see dashed circles in Fig. \ref{fig:Raw_data}(c)). Then, we take the inverse FFT of the windowed FFT to obtain a filtered measurement in real space, of which the absolute value is shown in Fig. \ref{fig:Raw_data}(e). Here, we observe that the background is greatly reduced and that the $\lambda_{SAW}$-periodic oscillations are removed. As a result, the signal emerging from the two slits is much more clearly visible. At first glance, the removal of the oscillations in the amplitude is surprising: by selecting the $\lambda_{SAW}$-component in the FFT of the \textit{complex} measurement, we have removed the $\lambda_{SAW}$- periodic oscillations from the \textit{amplitude} measurement. This removal is explained by the fact that the complex measurement allows us to separate the two interfering signals in the Fourier domain. As the background contribution has no spatial periodicity, its signal is located at the origin of Fig. \ref{fig:Raw_data}(c), whereas the SAW part is located on the aforementioned circle around $|k|=2\pi/\lambda_{SAW}$. Therefore, with our filtering technique, we removed the background signal and consequently the interference with the SAW signal, visible as oscillations in the amplitude measurement.

Furthermore, this filtering technique allows the observation of a much clearer interference pattern in the far-field, of which the fringes are bent slightly towards the positive $y$-direction. We also identify SAW propagation at high angles from the [110] axis and a small leakage signal, indicating that the grooves do not fully block the SAWs. In Fig. \ref{fig:Raw_data}(f), a close-up measurement of the region near the slits is shown. Here, we observe standing waves between the groove and the IDT (f.i) and transverse SAW confinement within the slits (f.ii) resulting in near-field interference effects in the positive $x$-direction (f.iii). We also observe circular diffraction from the groove edges (f.iv).

\begin{figure}[H]
\centering
\includegraphics[width=1\linewidth]{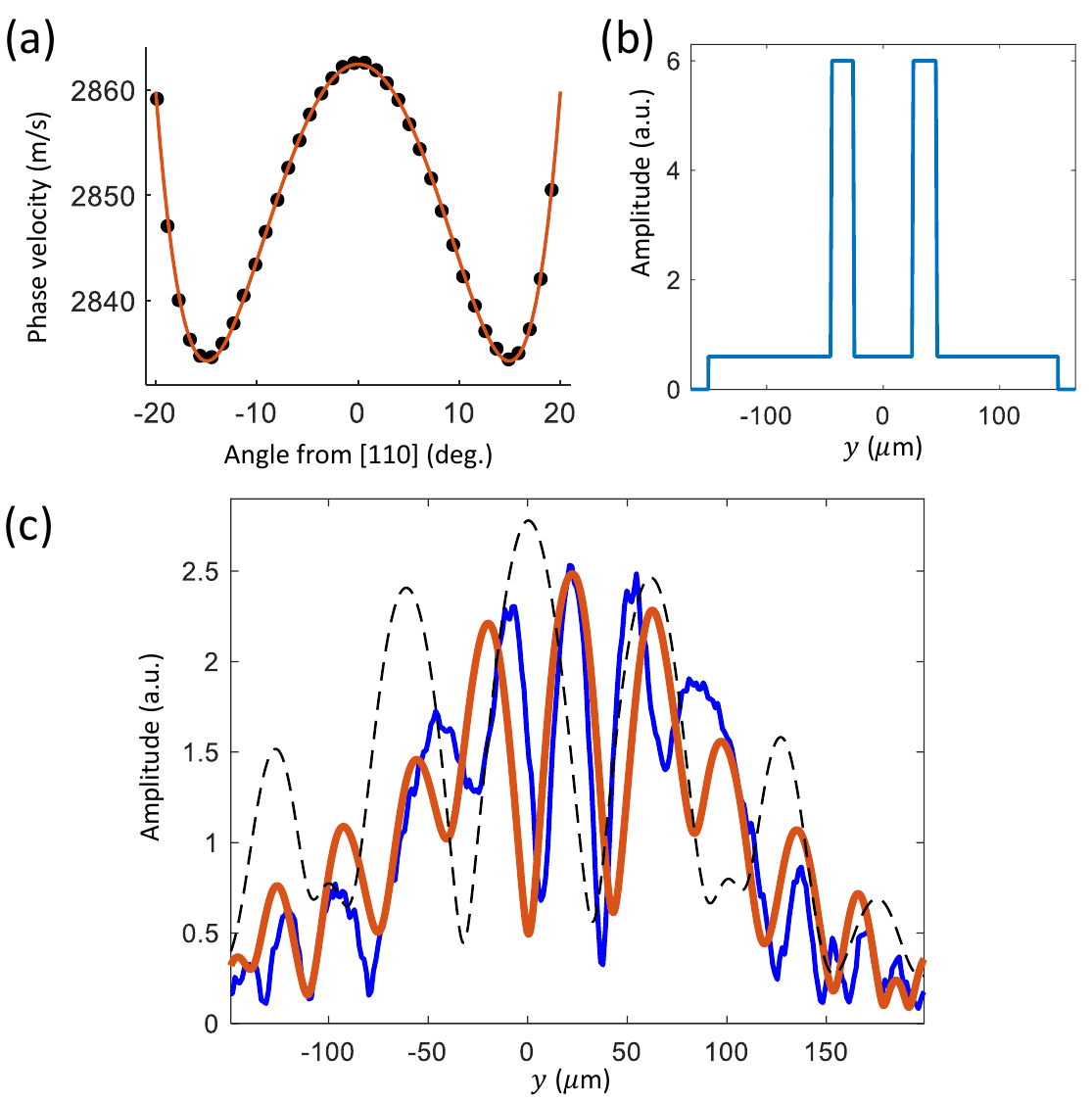}
\caption{Simulations. Angle-dependent SAW phase velocity (a) on GaAs. The black dots are extracted from Msall and Santos \cite{Msall2020} and the red curve is a polynomial fit used in the simulation. Input point sources $B_i$ at $x=0$ of the simulation (b) and comparison (c) between a slice of the filtered measurement at $x=$1500 $\mu$m (blue curve) and simulations with anisotropic phase velocity (red curve) and isotropic phase velocity (dashed black curve).}
\label{fig:Raw_filtmeas}
\end{figure}

Finally, we compare the experimental data to numerical simulations. For this, we use an anisotropic version of the Huygens-Fresnel principle to calculate the complex field $A(x,y)$ from $N$ point sources $B_i(x',y')$:   

\begin{equation}
    A(x,y)=\sum_{i=1}^{N}B_i(x',y')\frac{e^{i\omega r_i/v(\theta_i)}}{2r_i}(1+\cos{(\theta_i))}
\end{equation}

\noindent where $r_i$ and $\theta_i$ are given by:

\begin{equation}
\begin{aligned}
     r_i&=\sqrt{(x-x')^2+(y-y')^2} \\
    \theta_i&=\arctan{\bigg(\frac{y-y'}{x-x'}\bigg)}
\end{aligned}   
\end{equation}

\noindent and $\omega=2\pi f$, where $f=996 $ MHz is the RF generator frequency. The angle-dependent phase velocity $v(\theta)$ arises from the anisotropy of GaAs and is obtained by fitting of an even polynomial function to data provided by Msall and Santos \cite{Msall2020} and is valid up to $\pm$ 18$\degree$ with respect to the [110]-crystal axis, see Fig. \ref{fig:Raw_filtmeas}(a). The aforementioned tilt between the measurement frame and the [110]-crystal axis is taken into account by the rotation $v(\theta)=v(\theta - \phi)$. This tilt is chosen at 2.6$\degree$ such that the central peak of the simulation and measurement overlap. The input points of the simulation $B_i$ [see Fig. \ref{fig:Raw_filtmeas}(b)] are obtained from the slit geometry and include the leaked signal through the grooves. The relative amplitude between these signals is taken from the filtered measurement just after the slit. The simulation input contains no phase information and is set at $x=0$ of the measurement frame and thus only considers the field after the slits.

In Fig. \ref{fig:Raw_filtmeas}(c), a cross-section of the filtered measurement (at $x=$1500 $\mu$m) and two simulations are shown. The simulation with anisotropic SAW velocity (red curve) shows reasonable agreement with the measurement (blue curve) in terms of the periodicity of the interference fringes, as well as the relative amplitudes of the minima and maxima. We also performed a simulation without the leaked signal (not shown here), in which we obtained a near-unity fringe visibility. Therefore, we state that the leakage signal reduces the visibility of the interference fringes.

To study the effect of the anisotropy of the SAW velocity on the interference pattern, we also performed a simulation with an isotropic SAW phase velocity (dashed black curve in Fig. \ref{fig:Raw_filtmeas}(c)). The central interference peak of this simulation is not shifted in terms of $y-$ position, which indicates that due to the anisotropic wave velocity in combination with a tilt between the [110] crystal axis and the IDT and slit, the SAWs are steered in the direction of the [110] axis. This steering can not be explained by an angle between the IDT and the slit, as this angle is less than 0.1$\degree$. Secondly, the interference periodicity of the isotropic simulation is larger than in the measurement, indicating that the anisotropy decreases the interference periodicity. Both these findings are in agreement with theoretical predictions for 2D electron gasses in anisotropic media \cite{araujo}. We also would like to note here that the variation in $v(\theta)$ is relatively small (below 1\%), which means that the interference pattern in the far field is highly sensitive to small changes in $v(\theta)$.

We do observe, however, a small discrepancy between the anisotropic simulation (red curve) and the measurement (blue curve) in the periodicity of the interference fringes, see Fig. \ref{fig:Raw_filtmeas}(c). We hypothesize that this discrepancy might be related to slight variations in the SAW wavelength. Upon close inspection of the signal related to SAW propagation over the free surface in the FFT [(i) in Fig. \ref{fig:Raw_data}(d)], we find that this signal is slightly spread out in the $k_x$ direction and exhibits fringes. This could result in a discrepancy between the simulation and the measurement. A second potential origin of this discrepancy is the possible inaccuracy of the used angle-dependent velocity $v(\theta)$, which is based on a theoretical model \cite{Msall2020} and might be slightly different in reality. As the measurement technique presented here enables the extraction of the angle-dependent SAW wavelength, a direct measurement of $v(\theta)$ might be feasible with a dedicated sample where SAW propagation is directionally more uniform.

In conclusion, we present the first demonstration of Young's double-slit experiment with SAWs, where the slits are realized by milling grooves into the substrate with a FIB. Our measurement technique enables complete spatial characterization of the acoustic field, allowing us to reveal features that are difficult to capture in the optical counterpart of the experiment. After filtering in spatial frequency space, we found good agreement with anisotropic Huygens–Fresnel simulations. We find that accounting for the anisotropy of the SAW velocity on GaAs is essential, and we observe anisotropy-induced steering of the diffracted SAW beams.

\section*{Funding}
We acknowledge funding from NWO (QUAKE, No. 680.92.18.04), NWO/OCW (Quantum Software Consortium, Nos. 024.003.037, 024.003.037/3368), from the Dutch Ministry of Economic Affairs (Quantum Delta NL, No. NGF.1582.22.025), and from the European Union’s Horizon 2020 research and innovation program under Grant Agreement No. 862035 (QLUSTER).

\section*{Disclosures}
The authors declare no conflicts of interest.

\section*{Data availability}
Data underlying the results presented in this paper are not publicly available at this time but may be obtained from the authors upon reasonable request.

\bibliography{export}

\begin{thebibliography}{10}
\newcommand{\enquote}[1]{``#1''}

\bibitem{Young1804}
T.~Young, \enquote{The bakerian lecture. experiments and calculations relative to physical optics,} {\protect\JournalTitle{Philosophical Transactions of the Royal Society of London.}} \textbf{94}, 1--16 (1804).

\bibitem{Merli1976}
P.~G. Merli, G.~F. Missiroli, and G.~Pozzi, \enquote{On the statistical aspect of electron interference phenomena,} {\protect\JournalTitle{Am. J. Phys.}} \textbf{44}, 306--307 (1976).

\bibitem{Zeilinger1988}
A.~Zeilinger, R.~Gahler, C.~G. Shuii, \emph{et~al.}, \enquote{Single-and double-slit diffraction of neutrons,} {\protect\JournalTitle{Rev. Mod. Phys}} \textbf{60}, 1067--1073 (1988).

\bibitem{Arndt1999}
M.~Arndt, O.~Nairz, J.~Vos-Andreae, \emph{et~al.}, \enquote{Wave-particle duality of c 60 molecules,} {\protect\JournalTitle{Nature}} \textbf{401}, 680--682 (1999).

\bibitem{Fein2019}
Y.~Y. Fein, P.~Geyer, P.~Zwick, \emph{et~al.}, \enquote{Quantum superposition of molecules beyond 25 kda,} {\protect\JournalTitle{Nature Physics}} \textbf{15}, 1242--1245 (2019).

\bibitem{Andrews1997}
M.~Andrews, C.~Townsend, H.~Miesner, \emph{et~al.}, \enquote{Observation of interference between two bose condensates,} {\protect\JournalTitle{Science}} \textbf{275}, 637--641 (1997).

\bibitem{Zia2007}
R.~Zia and M.~L. Brongersma, \enquote{Surface plasmon polariton analogue to young's double-slit experiment,} {\protect\JournalTitle{Nature Nanotechnology}} \textbf{2}, 426--429 (2007).

\bibitem{Mezil2016}
S.~Mezil, K.~Chonan, P.~H. Otsuka, \emph{et~al.}, \enquote{Extraordinary transmission of gigahertz surface acoustic waves,} {\protect\JournalTitle{Scientific Reports}} \textbf{6} (2016).

\bibitem{Ash2021}
B.~J. Ash, A.~R. Rezk, L.~Y. Yeo, and G.~R. Nash, \enquote{Subwavelength confinement of propagating surface acoustic waves,} {\protect\JournalTitle{Applied Physics Letters}} \textbf{118} (2021).

\bibitem{Go2017}
D.~B. Go, M.~Z. Atashbar, Z.~Ramshani, and H.~C. Chang, \enquote{Surface acoustic wave devices for chemical sensing and microfluidics: A review and perspective,} {\protect\JournalTitle{Anal. Methods}} \textbf{9}, 4112--4134 (2017).

\bibitem{Rocha-Gaso2009}
M.~I. Rocha-Gaso, C.~March-Iborra, Ángel Montoya-Baides, and A.~Arnau-Vives, \enquote{Surface generated acoustic wave biosensors for the detection of pathogens: A review,} {\protect\JournalTitle{Sensors}} \textbf{9}, 5740--5769 (2009).

\bibitem{Sletten2019}
L.~R. Sletten, B.~A. Moores, J.~J. Viennot, and K.~W. Lehnert, \enquote{Resolving phonon fock states in a multimode cavity with a double-slit qubit,} {\protect\JournalTitle{Physical Review X}} \textbf{9} (2019).

\bibitem{Golter2016}
D.~A. Golter, T.~Oo, M.~Amezcua, \emph{et~al.}, \enquote{Coupling a surface acoustic wave to an electron spin in diamond via a dark state,} {\protect\JournalTitle{Physical Review X}} \textbf{6} (2016).

\bibitem{Patel2024}
S.~D. Patel, K.~Parto, M.~Choquer, \emph{et~al.}, \enquote{Surface acoustic wave cavity optomechanics with atomically thin h-bn and wse2 single-photon emitters,} {\protect\JournalTitle{PRX Quantum}} \textbf{5} (2024).

\bibitem{Decrescent2022}
R.~A. Decrescent, Z.~Wang, P.~Imany, \emph{et~al.}, \enquote{Large single-phonon optomechanical coupling between quantum dots and tightly confined surface acoustic waves in the quantum regime,} {\protect\JournalTitle{Physical Review Applied}} \textbf{18} (2022).

\bibitem{araujo}
F.~R. Araújo, D.~R.~D. Costa, F.~S.~D. Silva, \emph{et~al.}, \enquote{Single- and double-slit electron diffraction in an anisotropic two-dimensional medium,} {\protect\JournalTitle{Physical Review B}} \textbf{110} (2024).

\bibitem{Fisicaro2025}
M.~Fisicaro, T.~A. Steenbergen, Y.~C. Doedes, \emph{et~al.}, \enquote{Imaging transverse modes in a ghz surface acoustic wave cavity,} {\protect\JournalTitle{Phys. Rev. Applied}} \textbf{23}, 014032 (2025).

\bibitem{Msall2020}
M.~E. Msall and P.~V. Santos, \enquote{Focusing surface-acoustic-wave microcavities on gaas,} {\protect\JournalTitle{Physical Review Applied}} \textbf{13} (2020).

\end{thebibliography}
\bibliographyfullrefs{export}

\end{document}